\documentclass[a4paper,aps,prl,superscriptaddress,floatfix,nofootinbib,twocolumn,bibtex]{revtex4}

\usepackage{bbold}
\usepackage{bbm}
\usepackage[pdftex]{graphicx}
\usepackage{latexsym,amsmath,verbatim}
\usepackage{color}
\usepackage{rotating}
\usepackage{verbatim}
\usepackage{multirow}
\usepackage[english]{babel}
\usepackage{comment}

\begin{document}

\title{Observability transition in real networks}

\author{Yang Yang}
\affiliation{Department of Physics and Astronomy, Northwestern University, Evanston, Illinois 60208, USA}

\author{Filippo Radicchi}
\affiliation{Center for Complex Networks and Systems Research, School
  of Informatics and Computing, Indiana University, Bloomington,
  Indiana 47408, USA}
\email{filiradi@indiana.edu}

\begin{abstract}
 We consider the observability model
 in networks with arbitrary topologies. We introduce a system of 
 coupled nonlinear equations, valid under the 
 locally tree-like ansatz, to describe the
 size of the largest observable 
 cluster as a function
 of the fraction of directly observable nodes present in the network.
 We perform a systematic analysis on 95 real-world 
 graphs and compare our theoretical predictions with numerical 
simulations of the observability model.
Our method provides almost perfect predictions 
 in the majority of the cases, even 
 for networks with very large 
 values of the clustering coefficient.
Potential applications of our theory 
include the development of 
efficient and scalable algorithms 
for real-time surveillance of social networks, and
monitoring of technological networks.
\end{abstract}

\maketitle

The state of an entire networked dynamical system
can be determined by
monitoring or dominating  
the states of a limited 
number of nodes in the network ~\cite{liu2013observability}. 
A power-grid network can be 
observed in real time by placing 
phasor measurement units to a selection of nodes in the 
network~\cite{PhysRevLett.109.258701}.
Routing tables in mobile ad-hoc networks 
rely on gateway nodes to
form connected  dominating sets used as backbones for 
communication~\cite{wu1999calculating}.
Disease outbreaks in urban environments
can be efficiently detected by placing sensors 
on specific locations
visited by potentially infected
individuals~\cite{eubank2004modelling}. 

Whereas all these examples markedly differ in their 
underlying dynamics,
from the structural point of view, they can all be framed in terms of the
so-called network observability model~\cite{PhysRevLett.109.258701}.
In this model,  placing an observer on one node can 
make the node itself and all its nearest neighbors observable.
Nodes in the network can therefore assume three different
states: (i) directly observable, if hosting an observer; (ii) indirectly
observable, if being the first neighbor of an observer; (iii) or not
observable, otherwise. 
Observable, either directly 
or indirectly, nearest-neighbor nodes form clusters
of connected observable nodes. Thus, structurally speaking, 
the network observability model can be thought as
an extension of the more traditional,  and much more studied, 
percolation model~\cite{stauffer1991introduction, 
PhysRevLett.109.258701}. As in percolation, the 
question of  interest in network observability
is how to determine the macroscopic formation of observable clusters
in the network on the basis of microscopic changes
in the state of its individual nodes.

The observability model has been recently studied
in its simplest formulation where directly observed nodes
are randomly selected
~\cite{PhysRevLett.109.258701}. 
The model has been solved for both  
uncorrelated and correlated random network
models in the limit of infinite size~\cite{PhysRevLett.109.258701,
  PhysRevE.88.042809}.
As real networks are not mere realizations of
random network models, and their size
is clearly not infinite, the methods deployed in Refs.~\cite{PhysRevLett.109.258701,
  PhysRevE.88.042809} are not directly applicable
to real-world networks. The present paper
introduces a theoretical approach
able to describe the observability model in real graphs.
We introduce a set of heuristic equations that takes as
input the adjacency matrix of a network to draw its entire
observability phase diagram. The mathematical framework 
consists in the formulation of a belief-propagation
or message-passing algorithm~\cite{BF}
in a similar spirit as recent theoretical methods based
on message-passing algorithms have been used
to describe ordinary
percolation transitions in real isolated and/or interdependent 
networks~\cite{PhysRevLett.113.208701, 
PhysRevLett.113.208702,
RadicchiCastellano15, PhysRevE.93.030302, radicchi2015percolation}.
We show, through a systematic analysis of 
nearly one hundred
 real networks,
that the method is able to reproduce
true phase diagrams with extraordinary accuracy, 
proving therefore
its applicability to a wide range of real systems.

Here we consider an arbitrary 
network composed of $N$ nodes and $E$ edges.
Without loss of generality, we assume that the
network has
one single connected component. 
Suppose that each node has a probability  $\phi$ to host an observer, i.e. to be directly observable.
Nodes that are connected to directly observable nodes are, in turn, indirectly observable.
 Observable 
nearest-neighbor nodes form clusters.
For $\phi =0$, no nodes
are observable, hence
 there are no clusters. For $\phi =1$,
all nodes are directly observable, and thus they
form a single cluster.
At intermediate values of $\phi$, the network can be found
in two different phases:
(i)
 the regime of non-observability, 
where all clusters have microscopic size;
(ii)
 the phase of observability,
where a single macroscopic cluster,
comparable in size with the entire network,
is present. To monitor the transition
between these two phases, one usually relies on the order
parameter $P_\infty$, corresponding to the
relative size of the largest observable cluster (LOC). 
In the limit of infinitely large
networks, $P_\infty =0$, for $\phi \leq \phi_c$, and
$P_\infty > 0$, for $\phi > \phi_c$, 
with $\phi_c$ 
critical value of the
probability $\phi$. In the following, we describe a 
mathematical framework, deployed under the 
locally tree-like approximation, to
estimate the relative size of the LOC 
as a function of $\phi$.

\begin{figure}[t]
  \begin{center}
    \includegraphics[width=0.47\textwidth]{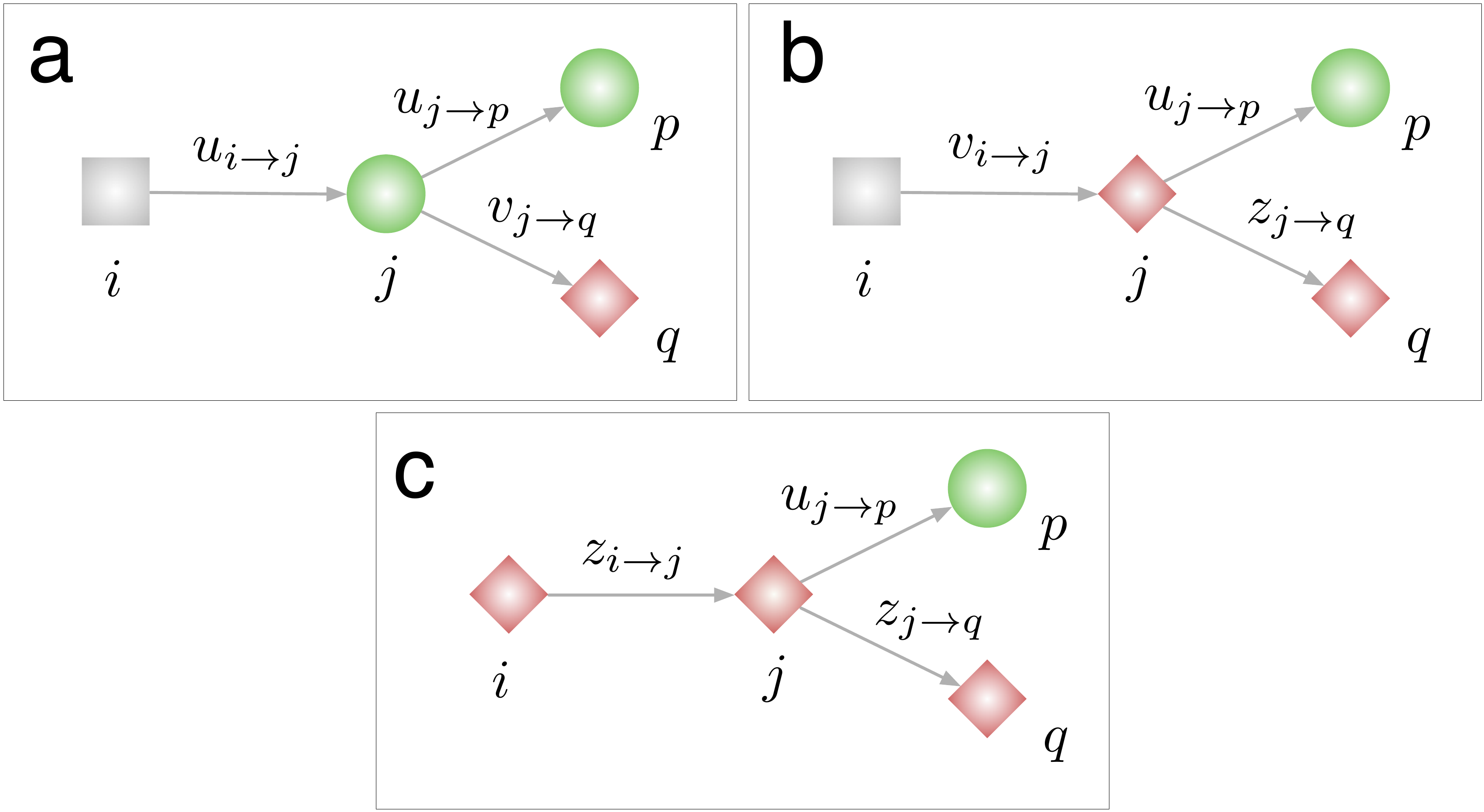}
  \end{center}
  \caption{
    Schematic illustration of the derivation of the system of 
    Eqs.~(\ref{eq:1}) [panel a],~(\ref{eq:2}) [b]
    and~(\ref{eq:3}) [c]. The different variables used in the equations
    are defined depending on the state of the nodes, here
    denoted by different shapes and colors (green circle = directly observable, 
    red diamond = not directly observable, and gray square = arbitrary). 
  }
\label{fig1}
\end{figure}

To proceed, we consider the probability 
that moving along 
the edge $i \to j$, we arrive to the LOC, irrespective 
of whether node $i$ is in the LOC or 
not~\footnote{Please note that the network is
undirected, but,
in our mathematical 
framework, every edge $(i, j)$
is considered twice, 
as $i \to j$ and $j \to i$.}. 
In particular, we consider three
conditional versions of this probability. 
We denote them as
$u_{i \to j}$ if $j$ is directly observable,
as $v_{i \to j}$ if $j$ is not directly observable, 
and as $z_{i \to j}$ if
neither $i$ nor $j$ are directly
observable.
Working under the
locally tree-like ansatz,  we can write the
following system of coupled equations (Fig.~\ref{fig1}):

\begin{equation}
  u_{i \to j} = 1 - \prod_{q \in \mathcal{N}_j \setminus \{i\} } [1 - \phi u_{j \to q} - (1 - \phi) v_{j \to q}] \; ,
  \label{eq:1}
\end{equation}

\begin{equation}
  v_{i \to j} = 1 - \prod_{q \in \mathcal{N}_j \setminus \{i\} } [1 - \phi u_{j \to q} - (1 - \phi) z_{j \to q}]
  \label{eq:2}
\end{equation}
and
\begin{equation}
  z_{i \to j} = v_{i \to j} - (1 - \phi)^{k_j-1} [ 1 - \prod_{q \in \mathcal{N}_j \setminus \{i\} } (1 - z_{j \to q}) ]  \; .
  \label{eq:3}
\end{equation}

In the above equations, $\mathcal{N}_j$
is the set of all neighbors of node $j$, and 
$k_j$ is the degree of node $j$. We note that $k_j = | \mathcal{N}_j
|$, where $| \mathcal{X}|$ indicates the size (i.e., number of
elements) of the set $\mathcal{X}$.
Equation~(\ref{eq:1}) is derived
as follows.
 If node $j$ is directly
observable, then node $j$ is 
part of the LOC
if at least one of its neighbors $q \neq i$ is part of the largest
cluster. This fact can happen in two ways: (i) with probability
$\phi \, u_{j \to q}$, if node $q$ is directly observable; (ii) with
probability $(1 - \phi) \, v_{j \to q}$, if node $q$ is not
directly observable. Thus, the probability 
that the connection $j \to q$ brings to the LOC is
$\phi \, u_{j \to q} + (1 - \phi) \, v_{j \to q}$.
The r.h.s. of Eq.~(\ref{eq:1}) quantifies the probability
that
at least 
one
of the connections $j \to q$
leads to the LOC, where the tree-like ansatz allows us to
consider probabilities associated with the individual
edges as independent variables, hence their product
appearing on the r.h.s. of Eq.~(\ref{eq:1}).

The derivation of Eq.~(\ref{eq:2}) is similar
to the one just described for Eq.~(\ref{eq:1}). We note
that we can write 
\begin{equation}
\begin{array}{l}
v_{i \to j} = 1 - \sum_{ \{s_r\}, r \in \mathcal{N}_j \setminus \{i \}
  }  
\\ \times \; 
\prod_{q
  \in \mathcal{N}_j \setminus \{i \}} \, [\phi (1 - u_{j \to q})
  ]^{s_q} \;  [(1 - \phi) (1 - z_{j \to q}) ]^{1 - s_q} 
\end{array}
\; .
\label{eq:4}
\end{equation}
For a  proof of the equivalence between Eqs.~(\ref{eq:2}) and~(\ref{eq:4}),
see SM. The 
sum on the r.h.s. of Eq.~(\ref{eq:4}) runs over all $2^{k_j-1}$ possible
configurations $\{s_r\}$ for the state (that is
directly  or not directly observable) of the neighbors
of node $j$, excluding node $i$.
For every given configuration,
the product appearing inside the
sum is the probability
that such a configuration appears, multiplied by 
the conditional probability that 
node $j$ is not attached to
the LOC in this configuration.
To be more specific, the binary variable $s_q=1$, if node $q$ is directly observable, and $s_q=0$, otherwise.
The quantity $[\phi (1 - u_{j \to q})
  ]^{s_q} [(1 - \phi) (1 - z_{j \to q}) ]^{1 - s_q}$ is the probability that the connection $j \to q$ does not bring node $j$
to the LOC.
Depending on whether node $q$ is directly observable or not, 
this probability is either $\phi (1 - u_{j \to q})$ or $(1 - \phi) (1 - z_{j \to q})$, respectively.

The expression of $z_{i \to j}$ in Eq.~(\ref{eq:3})
 can be quantified in almost the same way as $v_{i \to j}$.
We still need to consider the probabilities that the connection $i \to
j$ does not bring node $i$ to LOC, 
for all possible configurations of neighbors of node $j$.
The probability associated with each configuration
is the same as that appearing in  Eq.~(\ref{eq:4}).  
The only exception is the configuration $s_q = 0 \;, \forall q \in
\mathcal{N}_j \setminus \{i\}$,
which happens with probability $(1 - \phi)^{k_j-1}$,
where all neighbors of node $j$ are not directly observable 
(thanks to the underlying assumption that node $i$ is not directly
observable when we consider the conditional probability $z_{i \to  j}$), 
hence node $j$ is surely not observable and cannot be part of the LOC.
Accounting for this exception, and using
the equivalence between Eqs.~(\ref{eq:2}) and~(\ref{eq:4}),
we finally derive
Eq.~(\ref{eq:3}).

We can now rely on  
Eqs.~(\ref{eq:1}),~(\ref{eq:2}), and~(\ref{eq:3})
to compute the probability  $p_i$ that 
node $i$ is part of the LOC. 
We start from the simpler case when node $i$ is
directly observable, which happens
with probability $\phi$. 
We consider the probability 
that the connection $i \to j$ brings 
node $i$ to the LOC. 
This probability is $u_{i \to j}$, if node $j$ is directly observable,
and is $v_{i \to j}$, if node $j$ is not directly observable.
Combining the contributions from all neighbors of node $i$,
and using again the locally tree-like ansatz, the probability $r_i$ that node $i$ is 
directly observable, but not
part of the LOC is
\begin{equation}
  r_i = \phi \prod_{j \in \mathcal{N}_i} [1 - \phi u_{i \to j} - (1 -
  \phi)v_{i \to j}] \; .
\label{eq:r}
\end{equation}

 If node $i$
 is not directly observable, which happens
  with probability
$1 - \phi$, it is better to recast the
approach used to compute Eq.~(\ref{eq:3}).
We need to consider all possible $2^{k_i}$ configurations
for the neighbors of node $i$. Again, we have to account for
the special configuration $s_j = 0 \;, \forall j \in \mathcal{N}_i$,
when node $i$ is surely not observable. The probability $t_i$
that node $i$ is not directly observable, and none
of its neighbors is attached to the LOC
is given by
\[
\begin{array}{l}
t_i = (1 - \phi) \{ \sum_{ \{s_r\}, r \in \mathcal{N}_i}  \; 
\prod_{j
  \in \mathcal{N}_i }  [\phi (1 - u_{i \to j})
  ]^{s_j} 
\\
\times \, 
  [(1 - \phi) (1 - z_{i \to j}) ]^{1 - s_j} 
\\
+ (1 - \phi)^{k_i} - (1 - \phi)^{k_i} \prod_{j
  \in \mathcal{N}_i } (1 - z_{i \to j}) \} 
\end{array}
\; .
\]
Using the same trick 
as the one considered to pass from 
Eq.~(\ref{eq:4}) to Eq.~(\ref{eq:2}), we rewrite $t_i$ as
\begin{equation}
\begin{array}{l}
t_i = (1 - \phi) \{ \prod_{j
  \in \mathcal{N}_i } [1 - \phi u_{i \to j} - (1 - \phi)z_{i \to j}] 
\\
+ (1 - \phi)^{k_i} [ 1 - \prod_{j
  \in \mathcal{N}_i } (1 - z_{i \to j}) ] \} 
\end{array}
\; .
\label{eq:t}
\end{equation}
Combining the two cases, we derive the probability $p_i$ that node $i$ is part of the LOC as 
\begin{equation}
  \begin{array}{l}
    p_i =  1 - \phi \prod_{j \in \mathcal{N}_i} [ 1 - \phi u_{i \to j} - (1 - \phi)v_{i \to j}]
    \\
     - (1 - \phi) \{ \prod_{j \in \mathcal{N}_i} [1 - \phi u_{i \to j}
    - (1 - \phi)z_{i \to j}] 
\\
+ (1 - \phi)^{k_i} [1 - \prod_{j \in \mathcal{N}_i} (1 - z_{i \to j})]  \}
    \end{array} \; .
  \label{eq:5}
\end{equation}

The relative size of the LOC, predicted in the
locally tree-like ansatz, 
can be  finally calculated as 
\begin{equation}
P_\infty^{\text{(th)}} = \frac{1}{N} \sum_{i=1}^N p_i \; .
\label{eq:order}
\end{equation}
For every value of $\phi$, 
$P_\infty^{\text{(th)}}$ can be numerically estimated by
first solving by iteration  the system
of Eqs.~(\ref{eq:1}),~(\ref{eq:2})
and~(\ref{eq:3}) for every directed edge $i \to j$.
We can then plug the solution in the system
of Eqs.~(\ref{eq:5}), and estimate every $p_i$.
These values can be finally inserted in Eq.~(\ref{eq:order})
to compute $P_\infty^{\text{(th)}}$.

\begin{figure}[!htb]
  \begin{center}
    \includegraphics[width=0.47\textwidth]{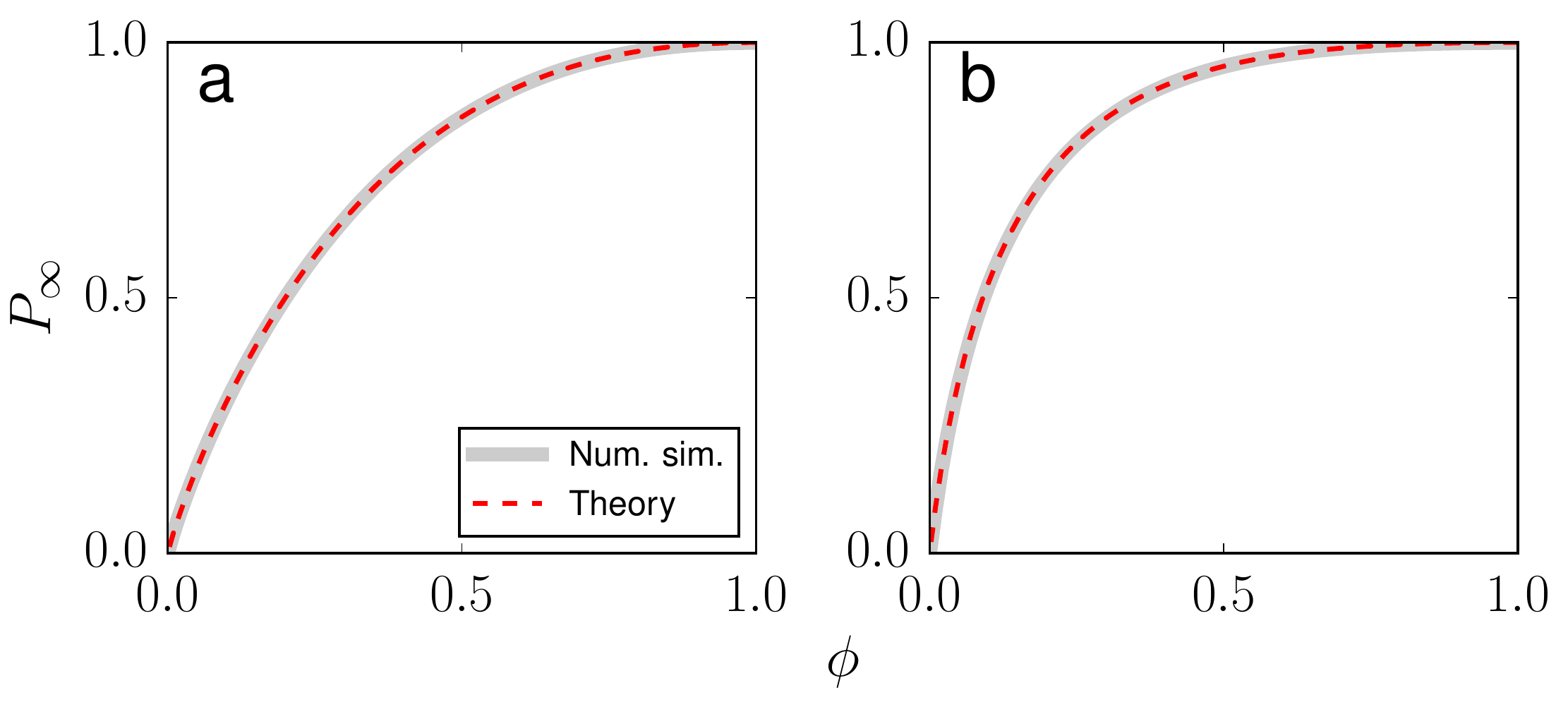}
  \end{center}
  \caption{Observability transition in real networks.
    We compare results from numerical simulations (gray lines)
    with the solution of our theoretical equations (red lines).
    (a) Analysis of the  Internet at the autonomous system
      level, as of July 22, 2006~\cite{PhysRevLett.113.208702}.
    (b) Analysis of the scientific collaboration network
    derived from pre-prints posted in the section Cond-Mat 
    of the arXiv between years 1993 and 2005~\cite{newman2001structure}.
  }
\label{fig2}
\end{figure}

In Fig.~\ref{fig2}, we present results from the analysis
of two real-world networks. The plots show a
comparison of the observability phase diagram
obtained from the solution of our framework, and the
one computed from numerical simulations of the model.
Simulations are performed using a modified version
of the Newman-Ziff algorithm, originally introduced
to simulate ordinary percolation processes in arbitrary 
topologies~\cite{newman2000efficient}. For every value
of $\phi$, we estimate the order parameter
$P_{\infty}^{\text{(num)}}$ as the average value over 
$10,000$ independent
realizations of the algorithm. 
The analysis of Fig.~\ref{fig2} reveals an almost 
perfect match between
theoretical predictions and results of numerical simulations.

\begin{figure}[!htb]
  \begin{center}
    \includegraphics[width=0.47\textwidth]{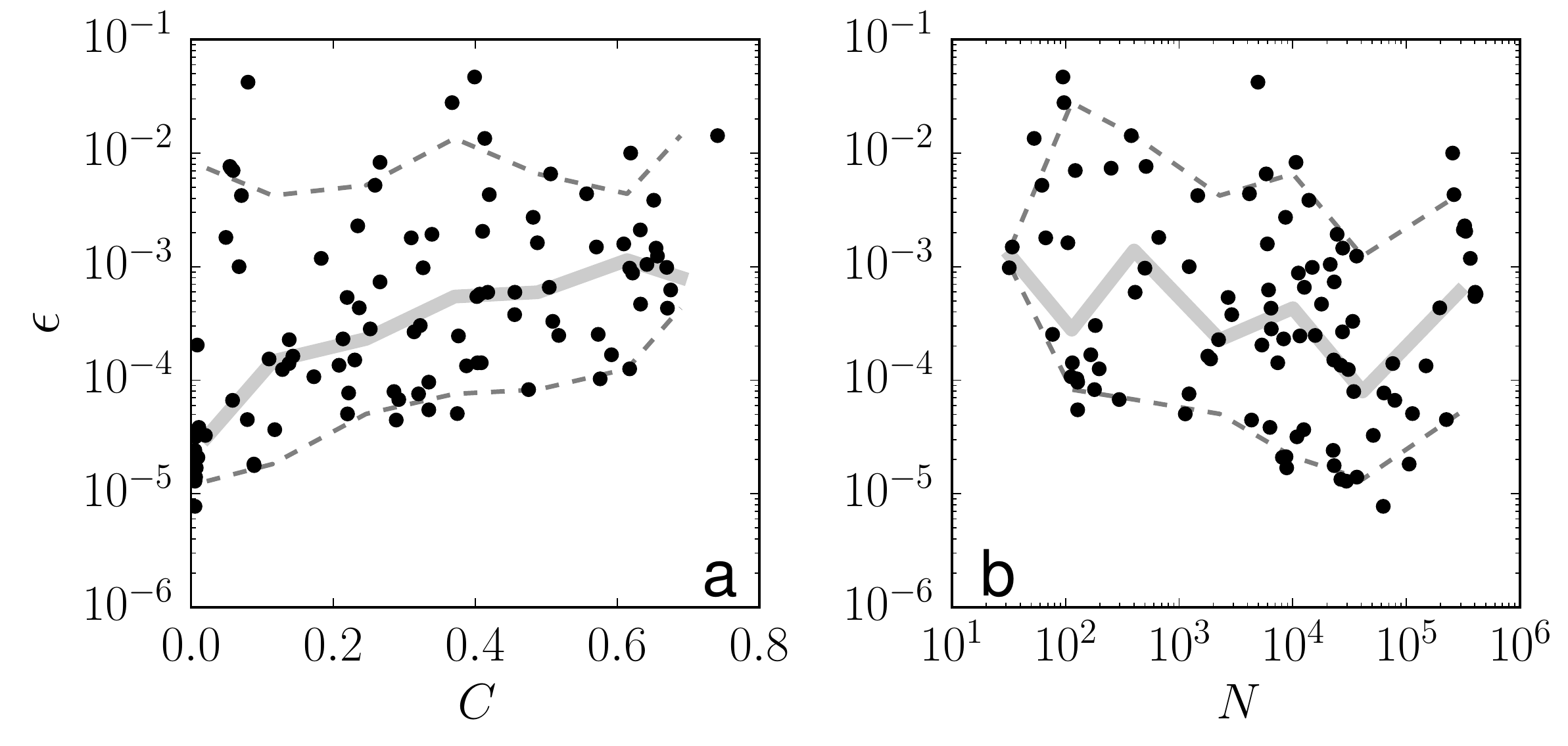}
  \end{center}
  \caption{
    Analysis of real networks.
    We consider $95$ real-world graphs~\cite{radicchi2014predicting}
    (see SM).
    For every network, we compute the discrepancy $\epsilon$ between 
    the theoretical and numerical estimates of the relative size of
    the LOC [Eq.~(\ref{eq:err})]. 
    (a) For every network,
    we plot $\epsilon$ as a function of the average clustering
    coefficient $C$. 
    To construct the lines, we consider seven
    equally spaced bins for the range of $C$ values. 
    For all networks falling in a given bin, we compute
    the median value of $\epsilon$ (full line), and
    the lower and upper 
    ends of
    $90\%$ confidence intervals
    (dashed lines). 
     (b) Scatter plot of
    $\epsilon$ 
    versus the network size $N$.
    Lines are constructed in a similar way as those appearing in
    panel a. The only difference is that we divide the range of $N$
    values in six equally spaced bins on the logarithmic scale.
  }
\label{fig3}
\end{figure}

To test how good our theoretical predictions are,
we perform a systematic comparison between
theory and numerical simulations
on $95$ real-world graphs~\cite{radicchi2014predicting}.
For a list of all networks analyzed see SM.
We consider networks of very different nature (e.g., technological,
social, biological), and
heterogeneous in terms
of their topological properties (e.g., clustering coefficient, 
size, degree distribution). 
To quantify the discrepancy between
theoretical
predictions and the ground truth offered by numerical simulations,
we use the following expression~\cite{Melnik11}
\begin{equation}
\epsilon = \int_{0}^{1} \, | P_{\infty}^{\text{(th)}} (\phi) -
P_{\infty}^{\text{(num)}} (\phi) | \; d\phi \; .
\label{eq:err}
\end{equation}

As the results of 
our analysis reveal, the
discrepancy between theory and numerical simulations 
is generally very small  (Fig.~\ref{fig3}). 
Besides,
we observe only a very 
weak
dependence
of $\epsilon$ on the average clustering coefficient of the network
$C$ (Fig.~\ref{fig3}a).
Because the theoretical framework is deployed under
the locally tree-like ansatz, and $C$ can be interpreted as a good
proxy for the degree of violation of this approximation, a positive correlation between $C$ and $\epsilon$
is expected. 
However, 
the error committed by our framework
to estimate the true observability diagram is very small
even for networks with extremely large values of the
clustering coefficient. This result is in stark contrast with 
what found for the ordinary site percolation model,
where high clustering implies large differences between
ground truth and approaches based on the
locally tree-like approximation~\cite{PhysRevE.93.030302}.
We further note that 
even for extremely small networks composed of tens of nodes,
theoretical predictions are very accurate. Moreover,
the discrepancy between theory and simulations
tends to decrease as the size of the network increases
(Fig.~\ref{fig3}b). This is also not a surprising result, given that our
theoretical framework is expected become exact in the limit
of (locally tree-like) infinite networks. 


Given the continuous nature of the observability phase transition,
in the vicinity of the critical point $\phi_c$, we can 
take a linear approximation of the system
of Eqs.~(\ref{eq:1}),~(\ref{eq:2})
and~(\ref{eq:3}), and rewrite them in matricial form as:
$\vec{u} = M [ \phi \vec{u} + (1 - \phi) \vec{v} ]$, 
$\vec{v} = M [ \phi \vec{u} + (1 - \phi) \vec{z} ]$, and
$\vec{z} = \vec{v} - R^{(\phi)}  \vec{z}$. 
In the 
above
expressions, $\vec{u}$, $\vec{v}$, and 
$\vec{z}$ are column vectors composed of $2E$ components, 
each corresponding to a directed edge of the graph.
Matrix
$M$ is the $2E \times 2E$ non-backtracking matrix 
of the graph, whose generic
element is defined 
as $M_{i \to j, \ell \to r} = \delta_{j, \ell} (1 - \delta_{i,r})$, 
with $\delta$ Kronecker symbol~\cite{hashimoto1989zeta, krzakala2013spectral}.
The generic element of the matrix $R^{(\phi)}$ is defined as
$R^{(\phi)}_{i \to j, \ell \to r} = (1 - \phi)^{k_j -1} \, M_{i
  \to j, \ell \to r}$.
Solving the previous system of linear
equations (see SM), we arrive to 
the eigenvalue/eigenvector equation
\begin{equation}
\vec{z} = \{ [ \mathbbm{1} -  M  \phi (1- \phi) (\mathbbm{1} - 
\phi M)^{-1}  M ]^{-1} \, (1- \phi) M  - R^{(\phi)} \}  \vec{z} \; .
\label{eq:phi_critical}
\end{equation}
Eq.~(\ref{eq:phi_critical}) serves to study the linear stability
of the trivial solution $\vec{z}^T = (0, \ldots, 0)$.
The critical value $\phi_c$ of the transition equals the
value of $\phi$ for
which the trivial solution becomes
unstable, and corresponds to the $\phi$ value for which
the operator appearing on the r.h.s. of Eq.~(\ref{eq:phi_critical})
has principal eigenvalue equal to one. 
Eq.~(\ref{eq:phi_critical}) is useful only in a limited number
of cases, as for example regular graphs (see SM).
For general networks instead,
solving Eq.~(\ref{eq:phi_critical})  is not computationally
efficient. This operation requires to determine 
the inverse of several matrices. 
From a numerical point of view, it is
thus better to rely on a binary search
combined with the numerical solution of the system 
of nonlinear 
Eqs.~(\ref{eq:1}),~(\ref{eq:2}), and~(\ref{eq:3}).
We further stress that the determination of the
critical point in the observability transition 
is not as meaningful
as in the case of percolation. The critical point $\phi_c$ is in fact
very close to zero for almost all networks. Thus, the 
emergence
of the LOC happens as soon as a very small number of
observers are randomly placed in the system.

Our  method to estimate observability 
phase diagrams is the first theoretical framework that 
can be applied to arbitrary network topologies.
Although
the method is 
exact 
only
for
locally tree-like infinite networks, its performances 
are almost perfect regardless of the size and/or
the average clustering coefficient  of the network.
In this paper, we considered and solved 
the ordinary version of the observability model, where
observers are randomly placed on nodes of the network.
We believe, however, that the framework 
has the potential to
be 
generalized to arbitrary strategies for the placement of observers.
In this sense, a very important extension of our formalism
could be to study optimal observability, on the same
footing as recent work on optimal
percolation~\cite{morone2015influence, dismantling, immuni}.
Considering that the optimal solution of the observability model 
is formally equivalent to the minimum (partial) dominating 
set of a graph~\cite{du2012connected}, 
such an extension could represent a very
important contribution for research in several domains, including, among others, 
biology~\cite{wuchty2014controllability, milenkovic2011dominating}
and social sciences~\cite{kelleher1988dominating, wang2011positive}.

\begin{acknowledgments}
The authors thank C. Castellano and A. Motter for critical comments 
on the early stages of this research. FR acknowledges 
support from the National Science 
Foundation (CMMI-1552487) and 
the US Army Research Office (W911NF-16-1-0104).
YY was supported by the National Science 
Foundation (DMS-1057128).
\end{acknowledgments}

\clearpage

\newpage
\onecolumngrid

\setcounter{page}{1}
\renewcommand{\theequation}{SM\arabic{equation}}
\setcounter{equation}{0}
\renewcommand{\thefigure}{SM\arabic{figure}}
\setcounter{figure}{0}
\renewcommand{\thetable}{SM\arabic{table}}
\setcounter{table}{0}

\section*{Appendix}

\subsection*{Proof by induction of Equation~(\ref{eq:4}) of the main text}

In the main text, we used the fact that
\begin{equation}
\prod_{q \in \mathcal{Q}} [1 - \phi a_q - (1 - \phi) b_q] =
\sum_{ \{s_r\}, r \in \mathcal{Q} }  \; 
\prod_{q
  \in \mathcal{Q}} \, [\phi (1 - a_q)
]^{s_q} \;  [(1 - \phi) (1 - b_q) ]^{1 - s_q} \; .
\label{eq:app1}
\end{equation}
Note that Eq.~(\ref{eq:app1}) is a more
general version of Eq.~(\ref{eq:4}).
We recall that sum on the r.h.s. runs over all
$2^{|\mathcal{Q}|}$ configurations, with $|\mathcal{Q}|$
number of elements in $\mathcal{Q}$, where
the element $q$ in the set $\mathcal{Q}$ can be in an active
state, i.e., $s_q=1$,
or in an inactive state, i.e., $s_q=0$.
These events happen with probability
$\phi$ and $1-\phi$, respectively,
providing the proper way to weight
the probability of appearance of
every configuration.
We provide here a proof by induction
of Eq.~(\ref{eq:app1}).
To this end, we first
note that if $\mathcal{Q} = \emptyset$, then
Eq.~(\ref{eq:app1}) is automatically satisfied, being both
sides equal to one. If $|\mathcal{Q}| > 0$,
we hypothesize that
\begin{equation}
\prod_{q \in \mathcal{Q} \setminus \{p\}} [1 - \phi a_q - (1 - \phi) b_q] =
\sum_{ \{s_r\}, r \in \mathcal{Q} \setminus \{p\} }  \; 
\prod_{q
  \in \mathcal{Q} \setminus \{p\}} \, [\phi (1 - a_q)
]^{s_q} \;  [(1 - \phi) (1 - b_q) ]^{1 - s_q} \;.
\label{eq:app2}
\end{equation}
The latter equation is the
analogue of Eq.~(\ref{eq:app1})
for the set
$\mathcal{Q} \setminus \{p\}$. We are thus supposing that
the equation is valid not for
the entire set $\mathcal{Q}$, but the set minus one its elements.
If we factorize out the contribution of
the element $p$ in Eq.~(\ref{eq:app1}), we have
\[
\begin{array}{l}
  [1 - \phi a_p - (1 - \phi) b_p] \;   \prod_{q \in \mathcal{Q} \setminus \{p\}} [1 - \phi a_q - (1 - \phi) b_q] =

  \\
  
  [ \phi (1- a_p) + (1 - \phi) (1 -b_p) ] \;  \sum_{ \{s_r\}, r \in \mathcal{Q} \setminus \{p\} }
\prod_{q
  \in \mathcal{Q} \setminus \{p\}} \, [\phi (1 - a_q)
]^{s_q} \;  [(1 - \phi) (1 - b_q) ]^{1 - s_q}
  \end{array} \; .
\]
By virtue of the hypothesis of Eq.~(\ref{eq:app2}),
the validity of Eq.~(\ref{eq:app1}) is obtained by
proving that the two extra factors due to the element $p$
that appear on both sides of the previous equation
are equal. This fact can be trivially shown
by rewriting
\[
\phi (1- a_p) + (1 - \phi) (1 -b_p) = 1 - \phi a_p - (1 - \phi) b_p \; .
\]

\subsection*{Linear approximation}

Using the the linear approximation
\[
\prod_q (1 - x_q) \simeq 1 - \sum_q x_q \; ,
\]

we can rewrite Eqs.~(\ref{eq:1}), ~(\ref{eq:2}) and ~(\ref{eq:3}) of
the main text
respectively as 

\begin{equation}
\vec{u} = M [ \phi \vec{u} + (1 - \phi) \vec{v} ] \; ,
\label{eq:1l}
\end{equation}

\begin{equation}
\vec{v} = M [ \phi \vec{u} + (1 - \phi) \vec{z} ]
\label{eq:2l}
\end{equation}
and 
\begin{equation}
\vec{z} = \vec{v} - R^{(\phi)}  \vec{z} \;, 
\label{eq:3l}
\end{equation}

where $R^{(\phi)}_{i \to j, \ell \to r} = (1 - \phi)^{k_j -1} \, M_{i
  \to j, \ell \to r}$, and 
$M$ is the non-backtracking matrix of the graph.

From Eq.~(\ref{eq:1l}), we obtain
\[
\vec{u} = (1- \phi) (\mathbbm{1} - \phi M)^{-1}  M \, \vec{v} \; .
\]

Inserting this expression into Eq.~(\ref{eq:2l}), we have
\[
\vec{v} =  [ \mathbbm{1} -  M  \phi (1- \phi) (\mathbbm{1} - \phi M)^{-1}  M ]^{-1} \, (1- \phi) M \, \vec{z} \; .
\]
Finally, using this expression in Eq.~(\ref{eq:3l}), we obtain
Eq.~(\ref{eq:phi_critical}).

A special case where Eq.~(\ref{eq:phi_critical})
can be simplified is for regular graphs with valency $k$, so that
$M$ and $R^{(\phi)}$ have the same eigenvectors.
We can write the condition for the critical probability $\phi_c$ as
\[
[1 - \phi_c (1- \phi_c) (1 - \phi_c \mu)^{-1}  \mu^2 ]^{-1} \, (1-
\phi_c) \mu  - (1-\phi_c)^{k-1} \mu = 1 \; ,
\]
with $\mu = k-1$.


\clearpage
\begin{table*}
\begin{center}\begin{tabular}{|l|r|r|r|r|r|} \hline
network & $N$ & $\epsilon$  &  $C$  &  Refs. & Url \\ 
 \hline 
{\tt Social 3} & $32$ & $0.000978$ & $0.3266$ & \cite{milo2004superfamilies} &\href{http://wws.weizmann.ac.il/mcb/UriAlon/index.php?q=download/collection-complex-networks}{url} \\ \hline 
{\tt Karate club} & $34$ & $0.001490$ & $0.5706$ & \cite{zachary1977information} &\href{http://www-personal.umich.edu/~mejn/netdata/}{url} \\ \hline 
{\tt Protein 2} & $53$ & $0.013476$ & $0.4135$ & \cite{milo2004superfamilies} &\href{http://wws.weizmann.ac.il/mcb/UriAlon/index.php?q=download/collection-complex-networks}{url} \\ \hline 
{\tt Dolphins} & $62$ & $0.005209$ & $0.2590$ & \cite{lusseau2003bottlenose} &\href{http://www-personal.umich.edu/~mejn/netdata/}{url} \\ \hline 
{\tt Social 1} & $67$ & $0.001794$ & $0.3099$ & \cite{milo2004superfamilies} &\href{http://wws.weizmann.ac.il/mcb/UriAlon/index.php?q=download/collection-complex-networks}{url} \\ \hline 
{\tt Les Miserables} & $77$ & $0.000253$ & $0.5731$ & \cite{knuth1993stanford} &\href{http://www-personal.umich.edu/~mejn/netdata/}{url} \\ \hline 
{\tt Protein 1} & $95$ & $0.046827$ & $0.3991$ & \cite{milo2004superfamilies} &\href{http://wws.weizmann.ac.il/mcb/UriAlon/index.php?q=download/collection-complex-networks}{url} \\ \hline 
{\tt E. Coli, transcription} & $97$ & $0.027824$ & $0.3675$ & \cite{mangan2003structure} &\href{http://wws.weizmann.ac.il/mcb/UriAlon/index.php?q=download/collection-complex-networks}{url} \\ \hline 
{\tt Political books} & $105$ & $0.001623$ & $0.4875$ & \cite{adamic2005political} &\href{http://www-personal.umich.edu/~mejn/netdata/}{url} \\ \hline 
{\tt David Copperfield} & $112$ & $0.000107$ & $0.1728$ & \cite{newman2006finding} &\href{http://www-personal.umich.edu/~mejn/netdata/}{url} \\ \hline 
{\tt College football} & $115$ & $0.000142$ & $0.4032$ & \cite{girvan2002community} &\href{http://www-personal.umich.edu/~mejn/netdata/}{url} \\ \hline 
{\tt S 208} & $122$ & $0.007029$ & $0.0591$ & \cite{milo2004superfamilies} &\href{http://wws.weizmann.ac.il/mcb/UriAlon/index.php?q=download/collection-complex-networks}{url} \\ \hline 
{\tt High school, 2011} & $126$ & $0.000103$ & $0.5759$ & \cite{fournet2014contact} &\href{http://www.sociopatterns.org/datasets/high-school-dynamic-contact-networks/}{url} \\ \hline 
{\tt Bay Wet} & $128$ & $0.000096$ & $0.3346$ & \cite{konect} &\href{http://konect.uni-koblenz.de/networks/foodweb-baywet}{url} \\ \hline 
{\tt Bay Dry} & $128$ & $0.000055$ & $0.3346$ & \cite{ulanowicz1998network, konect} &\href{http://konect.uni-koblenz.de/networks/foodweb-baydry}{url} \\ \hline 
{\tt Radoslaw Email} & $167$ & $0.000168$ & $0.5919$ & \cite{radoslaw, konect} &\href{http://konect.uni-koblenz.de/networks/radoslaw_email}{url} \\ \hline 
{\tt High school, 2012} & $180$ & $0.000083$ & $0.4752$ & \cite{fournet2014contact} &\href{http://www.sociopatterns.org/datasets/high-school-dynamic-contact-networks/}{url} \\ \hline 
{\tt Little Rock Lake} & $183$ & $0.000303$ & $0.3226$ & \cite{martinez1991artifacts, konect} &\href{http://konect.uni-koblenz.de/networks/maayan-foodweb}{url} \\ \hline 
{\tt Jazz} & $198$ & $0.000126$ & $0.6175$ & \cite{gleiser2003community} &\href{http://deim.urv.cat/~alexandre.arenas/data/welcome.htm}{url} \\ \hline 
{\tt S 420} & $252$ & $0.007366$ & $0.0561$ & \cite{milo2004superfamilies} &\href{http://wws.weizmann.ac.il/mcb/UriAlon/index.php?q=download/collection-complex-networks}{url} \\ \hline 
{\tt C. Elegans, neural} & $297$ & $0.000068$ & $0.2924$ & \cite{watts1998collective} &\href{http://www-personal.umich.edu/~mejn/netdata/}{url} \\ \hline 
{\tt Network Science} & $379$ & $0.014257$ & $0.7412$ & \cite{newman2006finding} &\href{http://www-personal.umich.edu/~mejn/netdata/}{url} \\ \hline 
{\tt Dublin} & $410$ & $0.000595$ & $0.4558$ & \cite{isella2011s, konect} &\href{http://konect.uni-koblenz.de/networks/sociopatterns-infectious}{url} \\ \hline 
{\tt US Air Trasportation} & $500$ & $0.000970$ & $0.6175$ & \cite{colizza2007reaction} &\href{https://sites.google.com/site/cxnets/usairtransportationnetwork}{url} \\ \hline 
{\tt S 838} & $512$ & $0.007634$ & $0.0547$ & \cite{milo2004superfamilies} &\href{http://wws.weizmann.ac.il/mcb/UriAlon/index.php?q=download/collection-complex-networks}{url} \\ \hline 
{\tt Yeast, transcription} & $662$ & $0.001809$ & $0.0490$ & \cite{milo2002network} &\href{http://wws.weizmann.ac.il/mcb/UriAlon/index.php?q=download/collection-complex-networks}{url} \\ \hline 
{\tt URV email} & $1,133$ & $0.000050$ & $0.2202$ & \cite{guimera2003self} &\href{http://deim.urv.cat/~alexandre.arenas/data/welcome.htm}{url} \\ \hline 
{\tt Political blogs} & $1,222$ & $0.000076$ & $0.3203$ & \cite{adamic2005political} &\href{http://www-personal.umich.edu/~mejn/netdata/}{url} \\ \hline 
{\tt Air traffic} & $1,226$ & $0.001000$ & $0.0675$ & \cite{konect} &\href{http://konect.uni-koblenz.de/networks/maayan-faa}{url} \\ \hline 
{\tt Yeast, protein} & $1,458$ & $0.004232$ & $0.0708$ & \cite{jeong2001lethality} &\href{http://www3.nd.edu/~networks/resources.htm}{url} \\ \hline 
{\tt Petster, hamster} & $1,788$ & $0.000163$ & $0.1433$ & \cite{konect} &\href{http://konect.uni-koblenz.de/networks/petster-friendships-hamster}{url} \\ \hline 
{\tt UC Irvine} & $1,893$ & $0.000154$ & $0.1097$ & \cite{opsahl2009clustering, konect} &\href{http://konect.uni-koblenz.de/networks/opsahl-ucsocial}{url} \\ \hline 
{\tt Yeast, protein} & $2,224$ & $0.000228$ & $0.1381$ & \cite{bu2003topological} &\href{http://vlado.fmf.uni-lj.si/pub/networks/data/bio/Yeast/Yeast.htm}{url} \\ \hline 
{\tt Japanese} & $2,698$ & $0.000536$ & $0.2196$ & \cite{milo2004superfamilies} &\href{http://wws.weizmann.ac.il/mcb/UriAlon/index.php?q=download/collection-complex-networks}{url} \\ \hline 
{\tt Open flights} & $2,905$ & $0.000378$ & $0.4555$ & \cite{opsahl2010node, konect} &\href{http://konect.uni-koblenz.de/networks/opsahl-openflights}{url} \\ \hline 
{\tt GR-QC, 1993-2003} & $4,158$ & $0.004387$ & $0.5569$ & \cite{leskovec2007graph} &\href{http://snap.stanford.edu/data/ca-GrQc.html}{url} \\ \hline 
{\tt Tennis} & $4,338$ & $0.000045$ & $0.2888$ & \cite{radicchi2011best} &\href{-}{url} \\ \hline 
{\tt US Power grid} & $4,941$ & $0.042156$ & $0.0801$ & \cite{watts1998collective} &\href{http://www-personal.umich.edu/~mejn/netdata/}{url} \\ \hline 
{\tt HT09} & $5,352$ & $0.000204$ & $0.0087$ & \cite{isella2011s} &\href{http://www.sociopatterns.org/datasets/hypertext-2009-dynamic-contact-network/}{url} \\ \hline 
{\tt Hep-Th, 1995-1999} & $5,835$ & $0.006561$ & $0.5062$ & \cite{newman2001structure} &\href{http://www-personal.umich.edu/~mejn/netdata/}{url} \\ \hline 
\end{tabular}
\end{center}
\caption{Summary table for real-world networks. The columns of the table respectively report: 
the name of the network, the number of nodes in the giant component, 
the value $\epsilon$ as defined in Eq.~(9) of the main text, 
the avearge clustering coefficient of the network,
reference(s) of the paper where the network has been first analyzed, 
and url of where the network data have been downloaded (to open the web page in your browser, just click on the word {\it url}).}
\label{table1}
\end{table*}

\clearpage
\begin{table*}
\begin{center}\begin{tabular}{|l|r|r|r|r|r|} \hline
network & $N$ & $\epsilon$  &  $C$  & Refs. & Url \\ 
 \hline 
{\tt Reactome} & $5,973$ & $0.001586$ & $0.6091$ & \cite{joshi2005reactome, konect} &\href{http://konect.uni-koblenz.de/networks/reactome}{url} \\ \hline 
{\tt Jung} & $6,120$ & $0.000623$ & $0.6752$ & \cite{vsubelj2012software, konect} &\href{http://konect.uni-koblenz.de/networks/subelj_jung-j}{url} \\ \hline 
{\tt Gnutella, Aug. 8, 2002} & $6,299$ & $0.000038$ & $0.0109$ & \cite{ripeanu2002mapping, leskovec2007graph} &\href{http://snap.stanford.edu/data/p2p-Gnutella08.html}{url} \\ \hline 
{\tt JDK} & $6,434$ & $0.000431$ & $0.6707$ & \cite{konect} &\href{http://konect.uni-koblenz.de/networks/subelj_jdk}{url} \\ \hline 
{\tt AS Oregon} & $6,474$ & $0.000283$ & $0.2522$ & \cite{leskovec2005graphs} &\href{http://snap.stanford.edu/data/as.html}{url} \\ \hline 
{\tt English} & $7,377$ & $0.000142$ & $0.4085$ & \cite{milo2004superfamilies} &\href{http://wws.weizmann.ac.il/mcb/UriAlon/index.php?q=download/collection-complex-networks}{url} \\ \hline 
{\tt Gnutella, Aug. 9, 2002} & $8,104$ & $0.000021$ & $0.0095$ & \cite{ripeanu2002mapping, leskovec2007graph} &\href{http://snap.stanford.edu/data/p2p-Gnutella09.html}{url} \\ \hline 
{\tt French} & $8,308$ & $0.000231$ & $0.2138$ & \cite{milo2004superfamilies} &\href{http://wws.weizmann.ac.il/mcb/UriAlon/index.php?q=download/collection-complex-networks}{url} \\ \hline 
{\tt Hep-Th, 1993-2003} & $8,638$ & $0.002722$ & $0.4816$ & \cite{leskovec2007graph} &\href{http://snap.stanford.edu/data/ca-HepTh.html}{url} \\ \hline 
{\tt Gnutella, Aug. 6, 2002} & $8,717$ & $0.000021$ & $0.0067$ & \cite{ripeanu2002mapping, leskovec2007graph} &\href{http://snap.stanford.edu/data/p2p-Gnutella06.html}{url} \\ \hline 
{\tt Gnutella, Aug. 5, 2002} & $8,842$ & $0.000017$ & $0.0072$ & \cite{ripeanu2002mapping, leskovec2007graph} &\href{http://snap.stanford.edu/data/p2p-Gnutella05.html}{url} \\ \hline 
{\tt PGP} & $10,680$ & $0.008295$ & $0.2659$ & \cite{boguna2004models} &\href{http://deim.urv.cat/~alexandre.arenas/data/welcome.htm}{url} \\ \hline 
{\tt Gnutella, August 4 2002} & $10,876$ & $0.000032$ & $0.0062$ & \cite{ripeanu2002mapping, leskovec2007graph} &\href{http://snap.stanford.edu/data/p2p-Gnutella04.html}{url} \\ \hline 
{\tt Hep-Ph, 1993-2003} & $11,204$ & $0.000879$ & $0.6216$ & \cite{leskovec2007graph} &\href{http://snap.stanford.edu/data/ca-HepPh.html}{url} \\ \hline 
{\tt Spanish} & $11,558$ & $0.000245$ & $0.3764$ & \cite{milo2004superfamilies} &\href{http://wws.weizmann.ac.il/mcb/UriAlon/index.php?q=download/collection-complex-networks}{url} \\ \hline 
{\tt DBLP, citations} & $12,495$ & $0.000037$ & $0.1178$ & \cite{ley2002dblp, konect} &\href{http://konect.uni-koblenz.de/networks/dblp-cite}{url} \\ \hline 
{\tt Spanish} & $12,643$ & $0.000658$ & $0.5042$ & \cite{konect} &\href{http://konect.uni-koblenz.de/networks/lasagne-spanishbook}{url} \\ \hline 
{\tt Cond-Mat, 1995-1999} & $13,861$ & $0.003846$ & $0.6514$ & \cite{newman2001structure} &\href{http://www-personal.umich.edu/~mejn/netdata/}{url} \\ \hline 
{\tt Astrophysics} & $14,845$ & $0.000986$ & $0.6696$ & \cite{newman2001structure} &\href{http://www-personal.umich.edu/~mejn/netdata/}{url} \\ \hline 
{\tt Google} & $15,763$ & $0.000248$ & $0.5176$ & \cite{palla2007directed} &\href{http://cfinder.org}{url} \\ \hline 
{\tt AstroPhys, 1993-2003} & $17,903$ & $0.000468$ & $0.6328$ & \cite{leskovec2007graph} &\href{http://snap.stanford.edu/data/ca-AstroPh.html}{url} \\ \hline 
{\tt Cond-Mat, 1993-2003} & $21,363$ & $0.001047$ & $0.6417$ & \cite{leskovec2007graph} &\href{http://snap.stanford.edu/data/ca-CondMat.html}{url} \\ \hline 
{\tt Gnutella, Aug. 25, 2002} & $22,663$ & $0.000024$ & $0.0053$ & \cite{ripeanu2002mapping, leskovec2007graph} &\href{http://snap.stanford.edu/data/p2p-Gnutella25.html}{url} \\ \hline 
{\tt Internet} & $22,963$ & $0.000151$ & $0.2304$ & - &\href{http://www-personal.umich.edu/~mejn/netdata/}{url} \\ \hline 
{\tt Thesaurus} & $23,132$ & $0.000018$ & $0.0888$ & \cite{kiss1973associative, konect} &\href{http://konect.uni-koblenz.de/networks/eat}{url} \\ \hline 
{\tt Cora} & $23,166$ & $0.000734$ & $0.2660$ & \cite{vsubelj2013model, konect} &\href{http://konect.uni-koblenz.de/networks/subelj_cora}{url} \\ \hline 
{\tt Linux, mailing list} & $24,567$ & $0.001928$ & $0.3391$ & \cite{konect} &\href{http://konect.uni-koblenz.de/networks/lkml-reply}{url} \\ \hline 
{\tt AS Caida} & $26,475$ & $0.000135$ & $0.2082$ & \cite{leskovec2005graphs} &\href{http://snap.stanford.edu/data/as-caida.html}{url} \\ \hline 
{\tt Gnutella, Aug. 24, 2002} & $26,498$ & $0.000013$ & $0.0055$ & \cite{ripeanu2002mapping, leskovec2007graph} &\href{http://snap.stanford.edu/data/p2p-Gnutella24.html}{url} \\ \hline 
{\tt Hep-Th, citations} & $27,400$ & $0.000267$ & $0.3139$ & \cite{leskovec2007graph, konect} &\href{http://konect.uni-koblenz.de/networks/cit-HepTh}{url} \\ \hline 
{\tt Cond-Mat, 1995-2003} & $27,519$ & $0.001457$ & $0.6546$ & \cite{newman2001structure} &\href{http://www-personal.umich.edu/~mejn/netdata/}{url} \\ \hline 
{\tt Digg} & $29,652$ & $0.000013$ & $0.0054$ & \cite{de2009social, konect} &\href{http://konect.uni-koblenz.de/networks/munmun_digg_reply}{url} \\ \hline 
{\tt Linux, soft.} & $30,817$ & $0.000124$ & $0.1286$ & \cite{konect} &\href{http://konect.uni-koblenz.de/networks/linux}{url} \\ \hline 
{\tt Enron} & $33,696$ & $0.000330$ & $0.5092$ & \cite{leskovec2009community} &\href{http://snap.stanford.edu/data/email-Enron.html}{url} \\ \hline 
{\tt Hep-Ph, citations} & $34,401$ & $0.000079$ & $0.2856$ & \cite{leskovec2007graph, konect} &\href{http://konect.uni-koblenz.de/networks/cit-HepPh}{url} \\ \hline 
{\tt Cond-Mat, 1995-2005} & $36,458$ & $0.001240$ & $0.6566$ & \cite{newman2001structure} &\href{http://www-personal.umich.edu/~mejn/netdata/}{url} \\ \hline 
{\tt Gnutella, Aug. 30, 2002} & $36,646$ & $0.000014$ & $0.0063$ & \cite{ripeanu2002mapping, leskovec2007graph} &\href{http://snap.stanford.edu/data/p2p-Gnutella30.html}{url} \\ \hline 
{\tt Slashdot} & $51,083$ & $0.000033$ & $0.0201$ & \cite{gomez2008statistical, konect} &\href{http://konect.uni-koblenz.de/networks/slashdot-threads}{url} \\ \hline 
{\tt Gnutella, Aug. 31, 2002} & $62,561$ & $0.000008$ & $0.0055$ & \cite{ripeanu2002mapping, leskovec2007graph} &\href{http://snap.stanford.edu/data/p2p-Gnutella31.html}{url} \\ \hline 
{\tt Facebook} & $63,392$ & $0.000077$ & $0.2218$ & \cite{viswanath2009evolution} &\href{http://socialnetworks.mpi-sws.org/data-wosn2009.html}{url} \\ \hline 
\end{tabular}
\end{center}
\caption{Continuation of Table~\ref{table1}.}
\label{table2}
\end{table*}

\clearpage
\begin{table*}
\begin{center}\begin{tabular}{|l|r|r|r|r|r|} \hline
network & $N$ & $\epsilon$  &  $C$  & Refs. & Url \\ 
 \hline 
{\tt Epinions} & $75,877$ & $0.000140$ & $0.1378$ & \cite{richardson2003trust, konect} &\href{http://konect.uni-koblenz.de/networks/soc-Epinions1}{url} \\ \hline 
{\tt Slashdot zoo} & $79,116$ & $0.000067$ & $0.0584$ & \cite{kunegis2009slashdot, konect} &\href{http://konect.uni-koblenz.de/networks/slashdot-zoo}{url} \\ \hline 
{\tt Flickr} & $105,722$ & $0.000018$ & $0.0884$ & \cite{McAuley2012, konect} &\href{http://konect.uni-koblenz.de/networks/flickrEdges}{url} \\ \hline 
{\tt Wikipedia, edits} & $113,123$ & $0.000051$ & $0.3748$ & \cite{brandes2010structural, konect} &\href{http://konect.uni-koblenz.de/networks/wikiconflict}{url} \\ \hline 
{\tt Petster, cats} & $148,826$ & $0.000134$ & $0.3877$ & \cite{konect} &\href{http://konect.uni-koblenz.de/networks/petster-friendships-cat}{url} \\ \hline 
{\tt Gowalla} & $196,591$ & $0.000434$ & $0.2367$ & \cite{cho2011friendship, konect} &\href{http://konect.uni-koblenz.de/networks/loc-gowalla_edges}{url} \\ \hline 
{\tt EU email} & $224,832$ & $0.000045$ & $0.0791$ & \cite{leskovec2007graph, konect} &\href{http://konect.uni-koblenz.de/networks/email-EuAll}{url} \\ \hline 
{\tt Web Stanford} & $255,265$ & $0.010009$ & $0.6189$ & \cite{leskovec2009community} &\href{http://snap.stanford.edu/data/web-Stanford.html}{url} \\ \hline 
{\tt Amazon, Mar. 2, 2003} & $262,111$ & $0.004310$ & $0.4198$ & \cite{leskovec2007dynamics} &\href{http://snap.stanford.edu/data/amazon0302.html}{url} \\ \hline 
{\tt DBLP, collaborations} & $317,080$ & $0.002107$ & $0.6324$ & \cite{ley2002dblp, konect} &\href{http://konect.uni-koblenz.de/networks/dblp_coauthor}{url} \\ \hline 
{\tt Web Notre Dame} & $325,729$ & $0.002287$ & $0.2346$ & \cite{albert1999internet} &\href{http://www3.nd.edu/~networks/resources.htm}{url} \\ \hline 
{\tt MathSciNet} & $332,689$ & $0.002048$ & $0.4104$ & \cite{palla2008fundamental} &\href{http://cfinder.org}{url} \\ \hline 
{\tt CiteSeer} & $365,154$ & $0.001183$ & $0.1832$ & \cite{bollacker1998citeseer, konect} &\href{http://konect.uni-koblenz.de/networks/citeseer}{url} \\ \hline 
{\tt Amazon, Mar. 12, 2003} & $400,727$ & $0.000545$ & $0.4022$ & \cite{leskovec2007dynamics} &\href{http://snap.stanford.edu/data/amazon0312.html}{url} \\ \hline 
{\tt Amazon, Jun. 6, 2003} & $403,364$ & $0.000595$ & $0.4177$ & \cite{leskovec2007dynamics} &\href{http://snap.stanford.edu/data/amazon0601.html}{url} \\ \hline 
{\tt Amazon, May 5, 2003} & $410,236$ & $0.000573$ & $0.4064$ & \cite{leskovec2007dynamics} &\href{http://snap.stanford.edu/data/amazon0505.html}{url} \\ \hline 
\end{tabular}
\end{center}
\caption{Continuation of Tables~\ref{table1} and~\ref{table2}.}
\label{tabl3}
\end{table*}

\end{document}